\documentclass[12pt]{iopart}

\usepackage{graphicx}
\usepackage{psfrag}
\usepackage{latexsym}
\usepackage{color}
\begin{document}

\title{Criticality and Condensation in  a Non-conserving Zero Range Process}
\author{A. G. Angel$^{1,2}$, M. R. Evans$^2$, E. Levine$^{3}$ and  D. Mukamel$^{4}$}

\address{
$^1$Center for Stochastic Processes in Science and Engineering,
Department of Physics,\\
 Virginia Tech, Blacksburg, VA 24061-0435, USA.\\
$^2$SUPA, School of Physics, University of Edinburgh, \\
Mayfield Road, Edinburgh EH9 3JZ, UK.\\
$^3$Center for Theoretical Biological Physics,\\
University of California at San-Diego, La Jolla, CA 92093, USA.\\
$^4$Department of Physics of Complex Systems, Weizmann Institute of Science, 
\\ Rehovot, Israel 76100.}
\ead{aangel@owl.phys.vt.edu,m.evans@ed.ac.uk, levine@ctbp.ucsd.edu, david.mukamel@weizmann.ac.il}

\begin{abstract}
The Zero-Range Process, in which particles hop between
sites on a lattice under conserving dynamics, is a 
prototypical model for studying real-space condensation.
Within this model the system is critical only at the transition point.
Here we consider  a non-conserving 
Zero-Range Process which is shown to exhibit generic critical phases
which exist in a range of creation and annihilation parameters.
The model also exhibits phases characterised by mesocondensates
each of which contains a subextensive number of particles.
A detailed phase diagram, delineating the various phases, is derived.
\end{abstract}


\date{\today}
\maketitle
\section{Introduction}
There exist many  systems exhibiting real-space
condensation under nonequilibrium steady state conditions.  
Examples
include jamming in traffic flow, granular clustering, wealth
condensation and gelation in networks \cite{EH05}.  Real-space condensation
implies that a finite fraction of some conserved quantity, for example
density, condenses onto a single lattice site or a small region in
space.  In general one is interested in the distribution of the
conserved quantity and the emergence of the condensate as some
parameter, often  the density, is varied.

The Zero-Range Process (ZRP) \cite{Spitzer} is a generic model which
exhibits condensation in its nonequilibrium steady state \cite{OEC98,Evans00}.  Moreover it
has the convenient property that its steady state is known and has a
simple factorised form. Thus the ZRP provides a simple exactly
solvable model within which general features of condensation may be
studied and analysed \cite{EH05}. For example, recent developments include:
extensions of the model to several conserved
quantities \cite{EH03,GS03,GLM05}, open boundary
conditions \cite{LMS05}, and sitewise
disorder \cite{Evans96,Krug00,JB03}; the study of current fluctuations
\cite{HRS05} and traffic modelling \cite{LZGM04,KMH05}.

The ZRP is usually defined on a lattice of $L$ sites where each site
may be occupied by any integer number of particles. The dynamics is
 defined by hopping rates $u(n)$ with which a particle hops from a
 site occupied by $n$ particles.  In various versions of the dynamics
 the particle may hop to different allowed destination sites. For
 example on a regular lattice particles hop to nearest neighbour sites whereas
 on a fully-connected geometry a particle can hop to any other site
 with equal probability. Clearly, the total number of particles, $N$, is
 conserved under the ZRP dynamics. Condensation occurs when in the large 
$N$, $L$ limit,
with the density $\rho = N/L$
held fixed, a finite
fraction of the particles condenses onto a single lattice site.
More recently,  an example where the condensate has
a non-zero spatial extent has been studied \cite{EHM06}.

Condensation is revealed in 
the single-site occupation probability distribution, $p(n)$.
A characteristic case  is when the hopping rate has the asymptotic, large $n$ form
$u(n) \simeq 1 + b/n$.
For $b>2$, the model exhibits a condensation transition at a critical density
$\rho_c$.
For $\rho < \rho_c$, $p(n)$
has the form
\begin{equation}
p(n) \propto n^{-b} {\rm e}^{-\mu n}
\label{p(n)}
\end{equation}
where $\mu$ is positive and is a function of the density.  Since there is
a characteristic occupation this is referred to as the fluid phase.

As the density increases towards the critical value $\rho_c$, $\mu$
tends to zero and the resulting distribution becomes a power law at
$\rho=\rho_c$.  For $\rho>\rho_c$ an extra piece of $p(n)$,
representing a single condensate, emerges centered at $n=L(\rho-\rho_c)$
\cite{EMZ05}.
Thus in the condensed phase a critical fluid co-exists with a
condensate containing the excess density.  A pure power-law
distribution only holds at the critical density which is a common
feature in many systems exhibiting phase transitions.  
For $b<2$ there is no condensation  transition since any density can be achieved
by choosing $\mu$ to be  suitably small in (\ref{p(n)}).

In this work we consider more general dynamical processes which could
allow for more complex condensation phenomena. Examples of the
phenomena we have in mind are the existence of criticality in an
extended region of the phase diagram rather than at an isolated point
(typically referred to as self-organised criticality),
and condensation into a large number of condensates each containing a
subextensive number of particles.  In our generalised dynamics we
introduce non-conserving processes with creation and annihilation
rates.  One mechanism to suppress a single extensive condensate  is
for the annihilation of particles to occur preferentially at highly
occupied sites.  We find that an appropriate choice of creation and
annihilation rates leads to a whole critical region of the phase
diagram where the occupation distribution decays algebraically at
large occupations.  Since the density is not conserved the phase
diagram is given in terms of the parameters of the creation and
annihilation rates.

Let us briefly summarise the phases which are exhibited (see
Fig.~\ref{phasediag}).  As might be expected, for imbalanced creation
and annihilation rates we find regimes with vanishing or diverging
density. These two regimes are separated by phases where the observed
density is equal to what would be the critical density on a conserving
system and power-law occupation distributions are exhibited. One of
these phases, Critical Phase A, exhibits a pure power-law distribution
for $p(n)$ with a cut-off diverging with system size.  However another
region in parameter space exists where the distribution exhibits, in
addition to the power-law decay, a broad and weak peak at high
occupations.  The height of the peak scales algebraically with system
size, rather than exponentially, as does the width.  This peak may
correspond to many `mesocondensates' each of which contains a
subextensive number of particles. The number of these mesocondensates
is also subextensive thus they occupy a vanishing fraction of the
sites.  We shall give a detailed analysis of this weak peak and how it
leads to two phases: Critical Phase B where the weak peak does not
contribute to the global density which is $\rho_c$ and the Weak High
Density Phase where the contribution of the weak peak to the global
density is dominant.

In a recent publication \cite{AELM05} we gave a brief account of the
non-conserving ZRP and its relation to self-organised criticality and
the dynamics of rewiring networks. Here we focus on the properties of
the non-conserving ZRP and present a detailed analysis of the phase
diagram.

The structure of the paper is as follows: in Section~\ref{sec:nczrp}
we define the non-conserving ZRP that we study; in
Section~\ref{sec:anal} we present a detailed analysis of the phase
diagram via a mean-field approximation and we compare the results to
numerical simulations on fully-connected and one-dimensional lattices;
we conclude with a discussion in Section~\ref{sec:disc}.

\section{Definition of non-conserving ZRP}
\label{sec:nczrp} We now define the non-conserving zero-range process which we
study in this work. 
The
lattice contains $L$ sites labelled $l = 1, \ldots, L$. The number of
particles at site $l$ is $n_l$.  
The dynamics are defined by the following three processes:
particles hop from a site with $n$ particles with rate
\begin{equation}
u(n) = \left(1 + \frac{b}{n}\right) \theta(n) \;  \label{uRate},
\end{equation}
where  the step function $\theta(n)$ is defined as
\begin{equation}
\theta(n) = \left\{ \begin{array}{c}
1 \quad \mbox{for}\quad n>0\;\\
0 \quad \mbox{for}\quad n=0\;,
\end{array}\right. 
\end{equation}
particles are created at a site with rate
\begin{equation}
c = \frac{1}{L^s} \; , \label{cRate}
\end{equation}
and particles evaporate from a site containing $n$ particles
with rate
\begin{equation}
a(n) = \left(\frac{n}{L}\right)^k \theta(n) \label{aRate}\; .
\end{equation}
In these rates the indices $k$ and $s$ are positive.
The creation rate at a site (\ref{cRate}) provides a weak drive which 
decreases with system size.
The particle annihilation rate at a site (\ref{aRate}) 
increases with the number of particles at the site and provides a mechanism
by which an extensive condensate may be suppressed.

As noted in the introduction, condensation in the conserving ZRP
occurs when $b>2$ and in the following we focus
primarily  on this
range of $b$.
For the specific  choice of hopping (\ref{uRate})
the critical density is known to be \cite{GSS03,Godreche03}
\begin{equation}
\rho_c = \frac{1}{b-2}\;.
\label{rhoc}
\end{equation}
Note that in the non-conserving case  $N$ and consequently
the density
$\rho$ fluctuate in time.

The precise nature of the lattice and the definition of the sites to
which particles are allowed to hop from a given site 
do not affect the qualitative picture emerging from this study
(as long as the lattice is homogeneous with all sites
having  the same hop rates).  To be specific we consider a
fully-connected lattice where a particle can hop to any other lattice
site.  This is a convenient  choice,
since on the fully-connected lattice the  mean-field approach with which we
treat the model analytically 
is expected to become exact in the large
system limit. In addition, we shall also present numerical data from one
dimensional systems with totally  asymmetric  nearest neighbour hopping
which indicate that the results of the mean-field analysis are relevant to this
case as well.

In a Monte-Carlo simulation the dynamics defined in
(\ref{uRate}--\ref{aRate}) are conveniently implemented as follows: At
each update
\begin{enumerate}
\item Select a site at random.
\item Generate a random number uniformly distributed between zero and the
sum of the maximum possible rates of the three processes, i.e.,
$c+\max(a(n))+\max(u(n))$.
\item If the random number falls in the range $(0,c)$ create a particle at
the site.
\item If the random number falls
in the range $(c,c+a(n))$, where
$n$ is the number of particles on the site,
evaporate a particle from the site.
\item If the random number falls in the range
$(c+\max(a(n)),c+\max(a(n)) + u(n))$ remove a particle from this site and
place it at another randomly selected site.
\end{enumerate}
Note, as
the evaporation rate $a(n)$ can diverge, a cut-off $\max(a(n))$ must be imposed
artificially and chosen large enough that, in practice,
the dynamics is not  affected.

The steady state of the model is fully described by the probability
distribution $P(n_1,n_2,\ldots,n_L)$ over all possible configurations.
In contrast to the conserving ZRP \cite{Evans00}, the steady-state
distribution of the model (\ref{uRate}--\ref{aRate}) does not factorize
generally.
However,  for the case of a fully-connected lattice we
expect factorisation to take effect  in the thermodynamic
limit  $L\to \infty$.
In the following analysis
we apply a  mean field approximation in which
the steady state distribution is replaced by a factorized form, i.e.
$P(n_1,n_2,\ldots,n_L) \to  \prod_{i=1}^L p(n_i)$.
For the  fully-connected lattice we
expect  this approximation
to become exact  in the thermodynamic
limit.

Within  this approximation, the  master equation is given
by
\begin{eqnarray}
\frac{\partial p(n)}{\partial t} &=& \left[ u(n+1) + a(n+1) \right]p(n+1)
- \left[ \lambda + c \right] p(n) \label{me1}\\
&&- \left\{ \left[ u(n) + a(n) \right]p(n) - \left[\lambda +
c\right]p(n-1)\right\} \theta(n)\;.  \nonumber
\end{eqnarray}
where $p(n)$ is the probability that a site contains $n$ particles.
The step function $\theta(n)$ ensures that (\ref{me1}) holds
for all $n \geq 0$.
The `current' $\lambda$ {\em into} a site due to the hopping  process  is given by
\begin{equation}
\lambda = \sum_{n=1}^{\infty} u(n) p(n)\;. \label{flux}
\end{equation}
To understand (\ref{me1}) note that $u(n) + a(n)$ is the total rate at which a site with $n$ particles loses
a particle and $\lambda + c$ is the total rate at which a site gains a particle.

In the steady state
(\ref{me1}) becomes
\begin{eqnarray}
0 &=& \left[ u(n+1) + a(n+1) \right]p(n+1)
- \left[ \lambda + c \right] p(n) \label{me2}\\
&&- \left\{ \left[ u(n) + a(n) \right]p(n) - \left[\lambda +
c\right]p(n-1)\right\} \theta(n).  \nonumber
\end{eqnarray}
which may be iterated to obtain
the steady-state  $p(n)$
\begin{equation}
p(n) = \frac{(\lambda + c)^n}{\prod_{m=1}^n[a(m) + u(m)]} p(0) \;.
\label{ssP}
\end{equation}
To fix $p(0)$ and $\lambda$, $p(n)$ must satisfy the following constraints of
normalisation and creation--annihilation balance respectively
\begin{eqnarray}
\sum_{n=0}^{\infty} p(n) &=& 1 \label{normalise}\\
\sum_{n=1}^{\infty} a(n) p(n) &=& c \label{balance} \; .
\end{eqnarray}
Only two of equations (\ref{flux}), (\ref{normalise}) and (\ref{balance})
are independent, as can be seen by noting
\begin{equation}
\left[ a(n) + u(n) \right] p(n) = (\lambda +c) p(n-1)\;,
\end{equation}
which, when summed, becomes
\begin{equation}
\sum_{n=1}^{\infty}\left[ a(n)+ u(n)\right] p(n)
=(\lambda +c) \sum_{n=0}^\infty p(n) =\lambda + c\;.
\end{equation}

\section{Analysis of phase diagram of non-conserving ZRP}
\label{sec:anal}
We now determine the asymptotic, large $L$ behaviour of
$\lambda$ (and therefore $p(n)$)
required to satisfy (\ref{flux}) and (\ref{balance}).
We identify  a total of five  phases in the $s$--$k$ plane: low density phase, 
strong and weak high density phases
and a critical region which itself consists of two distinct critical phases
as we shall explain  below.

\begin{figure}[htb]
\begin{center}
\includegraphics[width=7cm]{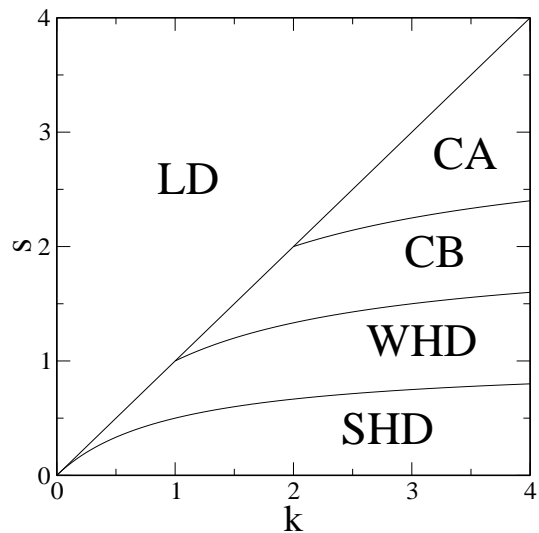}
\caption{\label{phasediag}
Typical phase diagram for the 
non-conserving ZRP model,  shown in the $k$--$s$ plane for
$b=3$.  The parameters $k,s,b$ are defined in 
(\ref{uRate}--\ref{aRate}). The labelling of the phases corresponds to
Low Density Phase (LD), Critical Phase A (CA), Critical Phase B (CB),
Weak High Density Phase (WHD) and Strong High Density Phase (SHD).}
\end{center}
\end{figure}

Identifying  the five  phases is most conveniently done
by inspecting the balance equation (\ref{balance}) which becomes
on inserting
the expressions (\ref{ssP}) for $p(n)$,  (\ref{cRate})
for $c$ and (\ref{aRate}) for $a$
\begin{eqnarray}
L^{k-s} &=& \sum_{n=1}^{\infty} n^k p(n) \nonumber \\
&=& p(0) \sum_{n=1}^{\infty} n^k
\exp\left[ n\,{\ln(\lambda+c)}
-\sum_{m=1}^n \ln(a(m) + u(m))\right] \label{genbal2}\;.
\end{eqnarray}

\subsection{Low density phase: $s>k$}
For $s > k$  the LHS of (\ref{genbal2})
tends to zero as the system size $L$ tends to infinity.  This
requires that $p(n)$ is a rapidly decreasing function of $n$.
To satisfy the balance equation (\ref{genbal2}) $\lambda + c$ has to be small.
Then to leading order in $L$, $\lambda +c \simeq \lambda \sim L^{k-s}$
\begin{equation}
p(n) \sim L^{-(s-k)n}\;.
\end{equation}
Thus the density $\rho \simeq p(1) \sim L^{k-s}$
which tends to zero as the system size tends to $\infty$.
The mean total number of particles $N$ increases sublinearly with
system size as $L^{k-s+1}$
and so we are in a low density phase when $s>k$.

\subsection{Analysis of $s<k$ region}

In this case, since
 $s < k$,  the LHS of (\ref{genbal2})
diverges with $L$, so the sum on the RHS must also diverge
and must be dominated by terms at large $n$.
This can only happen if $\lambda +c$ approaches one for large $L$.
Otherwise, if the limiting value of  $\lambda +c$ were less than one
the sum would not diverge, whereas if the limiting value were greater than one
the sum would diverge even for finite $L$.
Therefore we write
\begin{equation}
\lambda +c = 1 + h(L) \simeq e^{h(L)} \label{h(L)}\;,
\end{equation}
where $h(L)\to 0$ as $L\to \infty$.
Since we are interested in the large $n$ behaviour of $p(n)$,
keeping  leading order terms in $n$,
we have  $\ln(a(m)+u(m)) \simeq b/m + (m/L)^k$ and
then approximating  the sum by an integral gives, for large $n$,
$$\sum_{m=1}^n
\ln\left[ a(m)+u(m)\right] \simeq b\ln(n) + \frac{n^{k+1}}{(k+1)L^k}\;.$$
Therefore,  from (\ref{ssP}), the asymptotic form of $p(n)$
is 
\begin{equation}
p(n) \sim \frac{1}{n^b} \exp\left[  h(L) n  - \frac{n^{k+1}}{(k+1)L^k}\right]\;.
\label{pas}
\end{equation}
The rest of the analysis amounts to determining more precisely
the large $L$  behaviour of $h(L)$ required to satisfy the
balance equation (\ref{genbal2}) in the various regions of the $s$--$k$
plane.

\subsection{Strong High Density Phase: $s<k/(k+1)$}
The form of (\ref{pas}) suggests that if $h(L)$ is positive,
$p(n)$ may  be sharply peaked at some value $n^*$ which dominates
$p(n)$. By sharply peaked we mean that for $\epsilon$ arbitrarily small,
$\int_{n^*-\epsilon}^{n^*+\epsilon}
{\rm d}n\ p(n) \to 1$ as $L\to \infty$.
Then the balance equation (\ref{genbal2}) reduces to
$L^{k-s} \simeq (n^*)^k$
which implies
\begin{equation}
n^* \simeq L^{1-s/k}\;.
\label{nstar}
\end{equation}
This condition  may be used to determine $h(L)$.
Maximising the argument of the exponential  in (\ref{pas}) yields
$n^* = L h^{1/k}(L)$. Therefore
comparing with (\ref{nstar}) we require
\begin{equation}
h(L) \simeq L^{-s}\;.
\end{equation}
However, for the distribution to be sharply peaked at $n^*$
we require that the argument of the exponential in (\ref{pas}) diverges
 for large $L$ at $n^*$.
Thus, for example, $n^* h(L) \simeq L^{1-s-s/k}$ should diverge
which implies
\begin{equation}
s<\frac{k}{k+1}\;.
\end{equation}
In this phase, note that 
$\lambda +c = 1 + O(L^{-s})$
approaches one from above.
Also note that  the number of particles
is super extensive and  the mean density is $n^*$ which diverges as
(\ref{nstar}). The sharply peaked distribution $p(n)$ is rather different
from that of a conserving ZRP at high density, where  a critical fluid  and
a condensate piece coexist. In contrast, in the present case all sites contain
 $O(n^*)$ particles and we refer to this as the Strong High Density
Phase.

\subsection{Intermediate Regime}
We now consider the intermediate regime $k/(k+1)<s<k$
and show that $p(n)$ approaches a power-law distribution.
A careful analysis shows that, in fact,
this region may be divided into three  phases:
two of these are critical in the sense that the density
is given by $\rho_c$, the critical density of the conserving ZRP.
The first critical phase (Critical Phase A) has a power-law distribution $p(n)$
with a cut-off at large $n$. The second critical phase  (Critical Phase B)
has a power-law distribution together with a broad and weak peak at large 
occupations. In the thermodynamic limit this peak does not contribute to the global density and thus the global density is critical. Finally we have
a phase similar to Critical Phase B but where
the  broad peak dominates the global density
which now diverges with $L$. We refer to this as the Weak High Density Phase.

\subsubsection{Critical  Phase A $(kb/(k+1)\leq  s<k)$ }
\label{sec:cspa}
In the case  where $\lambda+c$ approaches one from below, namely when $h(L)$
in (\ref{h(L)})
is negative, the distribution (\ref{pas})
becomes a power law with a large $n$ cut-off
at $n \sim \mbox{min} [1/|h(L)|,  L^{k/(k+1)}]$. This cut-off
goes to infinity with $L$ resulting in a power-law distribution
in the thermodynamic limit.

We analyse the region in the $s$--$k$ plane where such a behaviour is
manifested.  Setting $h(L)\simeq - g L^{-x}$, where $g$ is a positive
constant, and replacing the sum by an integral,
the balance equation (\ref{genbal2}) takes the form
\begin{equation}
L^{k-s} \sim \int_1^{\infty} {\mathrm d}n\; n^{k-b}
\exp\left[-g\frac{n}{L^x}
-\frac{n^{k+1}}{(k+1)L^k} \right]\; . \label{powbal}
\end{equation}
In the case  $k>b-1$  the integral is
dominated by large $n$.
The cut-off in the large $n$ contribution
will be given by $L^x$, as long as  $x \leq k/(k+1)$.
The asymptotic behaviour of
(\ref{powbal}) becomes
\begin{eqnarray}
L^{k-s} &\sim& L^{x(k-b+1)}\;.
\end{eqnarray}
Thus, $x$ is given by
\begin{equation}
x = \frac{k-s}{k-b+1}\; .
\label{y}
\end{equation}
To be consistent with  the
requirement   $x\leq k/(k+1)$ one finds from (\ref{y})
that $s$ has to satisfy
\begin{eqnarray}
s &\geq& \frac{bk}{k+1}\;.
\end{eqnarray}
When the  equality holds, both terms in the exponential in
(\ref{powbal}) are relevant.
To summarise, in the range $kb/(k+1)\leq  s < k$ 
the asymptotic behaviour of $p(n)$ is a 
power law with an exponential cut-off,  the
cut-off point tending to infinity as the system size tends to
infinity:
\begin{equation}
p(n) \sim n^{-b}
\exp\left[-g \frac{n}{L^x}
\right]\; , \label{pascpb}
\end{equation}
where $x$ is given by (\ref{y}), and $g$ is an undetermined positive constant.
For this phase to exist we require $k>b-1$.

\subsubsection{The region $k/(k+1)\leq s<kb/(k+1)$}
\label{sec:cspb}
This still leaves the behaviour in the region $k/(k+1) < s < kb/(k+1)$
to be determined.  We expect that some kind of local maximum in $p(n)$
is needed in order to cope with the increased creation rate in this
region compared with Critical Phase A. However, a sharp peak in $p(n)$
is ruled out since that would correspond to the Strong High Density phase.
Instead we shall show that a `weak peak' emerges in $p(n)$ whose
height scales algebraically with $L$ as opposed to the sharp
exponential peak in the high density phase (see
Figure~\ref{weakpeak}).  We now demonstrate the existence of such a
peak in the distribution in the region $k/(k+1)<s<kb/(k+1)$, and show
how it satisfies the balance equation (\ref{genbal2}).

{
\psfrag{nstar}{$n^*$}
\psfrag{nmin}{$n_{\mathrm{min}}$}
\psfrag{pofnstar}{$p(n^*)$}
\psfrag{deltanstar}{$\Delta n^*$}
\begin{figure}[htb]
\begin{center}
\includegraphics[width=7cm]{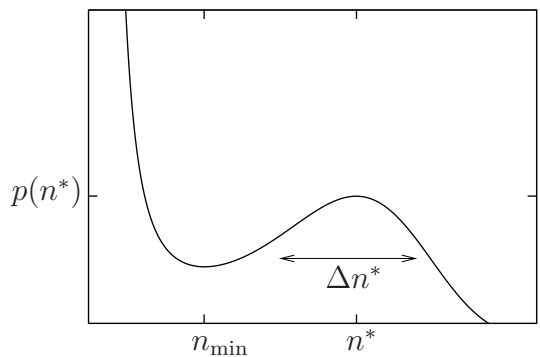}
\caption{\label{weakpeak} Schematic illustration
of a weak peak in $p(n)$.
Note that both the height $p(n^*)$ and width $\Delta n^*$
scale algebraically with system size $L$ but the weight.
$w = \Delta n^* p(n^*)$ of the peak
vanishes. The dip in the distribution
to the left of the weak peak is denoted $n_{\rm min}$.}
\end{center}
\end{figure}
}

Let us write the distribution (\ref{pas})  as
\begin{equation}
p(n) \sim e^{\psi(n)}
\end{equation}
where
\begin{equation}
\psi(n) = h(L)n - \frac{1}{k+1}\frac{n^{k+1}}{L^k} -b \ln n\;.
\label{psi}
\end{equation}
We now seek a function  $h(L)$ which produces a maximum at $n^*$
and for which $\psi(n^*) \sim \ln L$. This would result in
$p(n^*)$ scaling as a power of $L$.
In order for the second term in  (\ref{psi})
to scale as $\ln L$ we require
$n^* \sim (L^k \ln L )^{ 1/(k+1)}$. Then for the first term to scale similarly
we require
\begin{equation}
h(L) =
\left(\frac{d\ln L}{L} \right)^{k/(k+1)}\;,
\end{equation}
where $d$ is some constant to be determined. Having deduced the required
functional form of $h(L)$ we now turn to determining the constant $d$ and the precise
location and properties of the weak peak.

Taking the first two derivatives of $\psi(n)$  we have
\begin{eqnarray}
\psi^{\prime}(n) &=&
\left( \frac{d\ln L}{L} \right)^{k/(k+1)}
- \frac{n^k}{L^k}  -\frac{b}{n} \label{psi'}\\
\psi^{\prime\prime}(n) &=& - k\frac{n^{k-1}}{L^k} +\frac{b}{n^2}\;.
\end{eqnarray}
Setting $\psi'(n^*)=0$, gives to leading order in $L$
\begin{eqnarray}
n^{*} &\simeq & d^{1/(k+1)}  L^{k/(k+1)}
\left( \ln L  \right)^{1/(k+1)} \label{nstarwp}\;, \\
\psi(n^{*}) &\simeq& \frac{(d-b)k}{k+1}\ln L  -\frac{b}{k+1} \ln \ln L\;,\\
\psi^{\prime\prime}(n^{*}) &\simeq& -kL^{-2k/(k+1)}
\left( \ln L  \right)^{(k-1)/(k+1)}  d^{(k-1)/(k+1)} \;.
\end{eqnarray}
These expressions  imply that the height
of the peak at $n^{*}$  is
\begin{equation}
 p(n^*)\sim L^{k(d-b)/(k+1)}
\left(\ln L \right)^{-b/(k+1)}\;,
\label{height}
\end{equation}
and the width of the peak is given by
\begin{equation}
\Delta n^* = \left|\psi^{\prime\prime}(n^{*})\right|^{-1/2} \sim
L^{k/(k+1)}\left(\ln L \right)^{-(k-1)/(2(k+1))}\;.
\end{equation}
Thus the weight $w$ of the weak peak  in $p(n)$,
which we define as
\begin{equation}
w = p(n^*) \Delta n^*
\label{wpw}
\end{equation}
is given by
\begin{equation}
w \sim L^{(d-b+1)k/(k+1)} (\ln L)^{-(k-1+2b)/2(k+1)}\;.
\label{w}
\end{equation}

The balance equation (\ref{genbal2}) is satisfied by the contribution of the weak peak
and becomes asymptotically, 
\begin{equation}
L^{k-s} \sim (n^*)^k w\;.
\end{equation}
Inserting the expressions (\ref{w}) and (\ref{nstarwp}) implies,
ignoring logarithmic factors,
\begin{equation}
d = b - \frac{s\,(k+1)}{k}\;.
\label{d}
\end{equation}
Thus, from (\ref{pas}), $p(n)$  behaves asymptotically as
\begin{equation}
  p(n) \sim n^{-b} \exp \left[ n \left\{
\left( b - \frac{s(k+1)}{k} \right)
 \frac{\ln L}{L} \right\}^{k/(k+1)} - \frac{n^{k+1}}{(k+1)L^k} \right] \;.
\label{cspbpn}
\end{equation}
When $s<kb/(k+1)$ one has $d>0$ as required.
Also we require that the weight
(\ref{w}) of the
peak should not diverge as $L \to \infty$ otherwise  the system
would be in the Strong High Density Phase. This implies from (\ref{w}) and (\ref{d})
that $s \geq  k/(k+1)$.

To summarise, for $k/(k+1)\leq s<kb/(k+1)$ we have a probability distribution which initially  decreases from a
finite $p(0)$ as a power law as in Critical Phase A, but with a weak peak  
at high
$n$. The weak peak  allows the creation-annihilation balance condition to be
satisfied (see Fig~\ref{weakpeak}).
At this point let us recap
the properties of the weak peak, ignoring logarithmic correction factors:
\begin{eqnarray}
n^* &\sim& \Delta n^* \sim L^{k/(k+1)}\label{wp1}\\
p(n^*)&\sim& L^{-s}\\
w &\sim& L^{k/(k+1)-s}\;. \label{wp3}
\end{eqnarray}

Within the region
$k/(k+1)\leq s<kb/(k+1)$ there are in fact two phases distinguished by
the behaviour of the global density.
To see this we note that
the number of particles per site, $n_{wp}$, associated with the
weak peak is
\begin{equation}
n_{wp}\simeq n^* w \sim L^{2k/(k+1)-s}
\end{equation}
This gives the contribution of the weak peak to the density.
Clearly this contribution diverges or goes to zero according to whether
or not $s<2k/(k+1)$, thus implying two distinct regimes.

\noindent {\bf Critical Phase B}:  $2k/(k+1)<s<kb/(k+1)$\\
In this region $n_{wp}$ approaches zero in the large $L$
limit and the global density of the system is controlled by the power law
part of $p(n)$. Thus
the global density is $\rho=\rho_c$ where $\rho_c$ is the critical
density of the corresponding conserving ZRP (\ref{rhoc}).

\noindent {\bf Weak High Density Phase}: $k/(k+1)\leq s<2k/(k+1)$\\
In this region $n_{wp}$ diverges
so the density is dominated by the weak peak
and the total number of particles increases as
$N \sim L^{2k/(k+1) -s+1}$.

In these two phases
most of the sites of the system form a fluid
with typically low occupation numbers. In addition a subextensive
number of sites, $L w \sim L^{1-s+k/(k+1)}$ are highly occupied with
$n \simeq n^*$ particles where $n^* \sim L^{k/(k+1)}$ diverges
sublinearly with $L$.  We term these highly occupied sites
{\it  mesocondensates}.  The expected number of such mesocondensates is given
by $wL \sim L^{1-s+k/(k+1)}$
which can vary 
from $O(L)$ to $O(L^{1-k(b-1)/(k+1)})$.  If
\begin{equation}
b > 2 + \frac{1}{k}\;,
\end{equation}
 the average number of mesocondensates can be very much less than one.
Thus, in this case we  typically  do not expect to observe any mesocondensates.
In terms of
$s$, this happens when
\begin{equation}
\frac{kb}{k+1} > s > \frac{2k + 1}{k+1}\;.
\end{equation}

Another distinction between
Critical Phase A and the phases characterised by a weak peak
is the behaviour of  arbitrary moments of $n$
\begin{equation}
\mu_r = \int {\rm d}n\, n^r p(n)\;.
\end{equation}
In  Critical Phase A $\mu_r$ diverges when $r > b-1$.
On the other hand, the contribution  to $\mu_r$
from a  weak peak is $(n^*)^r w \sim L^{(r+1)/(k+1) -s}$
which diverges when $r> s (k+1)/k -1$.  Therefore
in Critical Phase B and the Weak High Density Phase
which moments diverge depends on 
the precise values of $s,k$. This is in contrast to 
Critical Phase A
where throughout the phase the same moments ($\mu_r$ for $r> b-1$) diverge.

\subsection{Summary of phase diagram}
\begin{figure}[htb]
\begin{center}
\includegraphics[width=10cm]{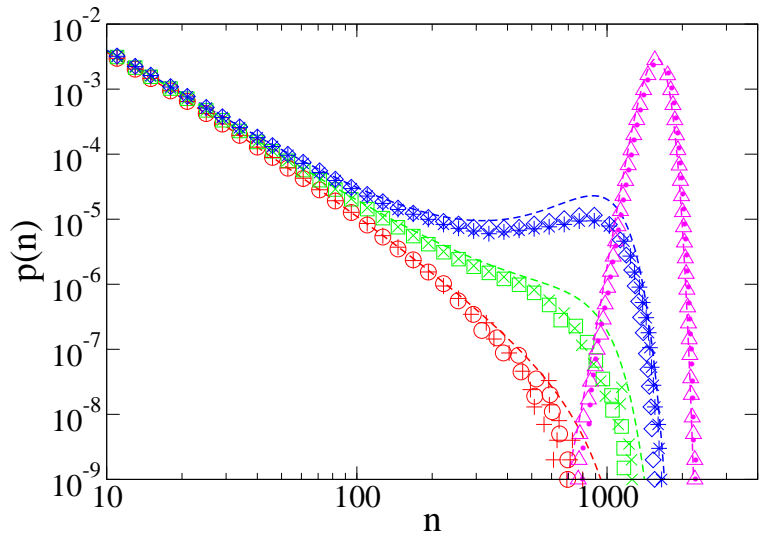}
\caption{\label{phasefig}
Steady-state distributions of the number of particles on a site from
simulations of the non-conserving ZRP model on a fully connected lattice
(open shapes) and a 1d lattice (character symbols), compared with
theoretical curves (dashed lines).  Simulations were run on a system with
L=5000 lattice sites and $b=2.6$, $k=3$.  Data are shown for:
Critical Phase A, $s=2$
(${\color{red}\circ}$, ${\color{red}+}$ and {\color{red}- -});
Critical Phase B, $s=1.7$
(${\color{green}\Box}$, ${\color{green}\times}$ and {\color{green}- -});
Weak High Density Phase, $s=1.2$
(${\color{blue}\Diamond}$, ${\color{blue}*}$ and {\color{blue}- -});
Strong High Density Phase, $s=0.4$
(${\color{magenta}\triangle}$, ${\color{magenta}\bullet}$ and
{\color{magenta}- -}).}
\end{center}
\end{figure}

\begin{figure}[t]
\begin{center}
\includegraphics[width=10.0cm]{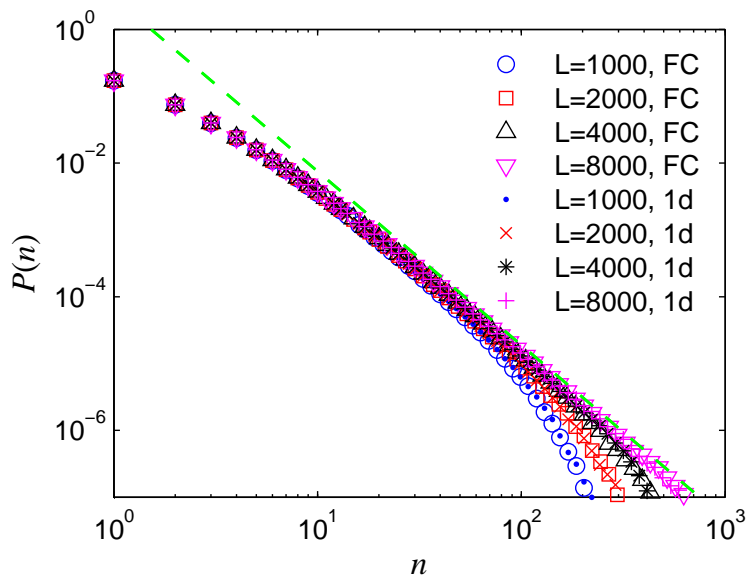}
\caption{\label{phasefig2}
Steady state probability distributions from simulations
in  Critical Phase A, illustrating $L$
dependence of the cut-off.
Here $b=2.6$, $k=3$ and $s=2$,
with $L=1000$, 2000, 4000, 8000.
The  legend indicates for each data set the 
system size and the geometry used:  fully-connected lattice
(FC) or one-dimensional lattice (1d). The dashed line corresponds to $P(n) \sim  n^{-b}$. }
\end{center}
\end{figure}

We have identified the following phases which are illustrated in 
a typical phase diagram in Figure~\ref{phasediag}
\begin{itemize}
\item $k<s$ --- {\bf Low Density Phase} (LD)

Here $p(n)$ decays as $L^{-(s-k)n}$
and the global density vanishes as $L^{-(s-k)}$.
\item $kb/(k+1)\leq s <k$ --- {\bf Critical  Phase A} (CA)

Here $p(n)$ decays algebraically as $1/n^b$
with a cut-off at $L^y$ where  $y=(k-s)/(k-b+1)$.
The global density throughout this phase is the critical density $\rho_c$.

\item $2k/(k+1)\leq s<bk/(k+1)$ --- {\bf Critical  Phase B} (CB)

Here $p(n)$ decays algebraically as $1/n^b$
and is cut-off by a weak peak, with properties (\ref{wp1}--\ref{wp3}),
  at $n^* \sim (\ln L)^{1/(k+1)} L^{k/(k+1)}$.
The global density is the critical density $\rho_c$.

\item $k/(k+1)\leq s<2k/(k+1))$ --- {\bf Weak High Density Phase}  (WHD)

Here $p(n)$ decays algebraically as $1/n^b$
and is cut-off by a weak peak  at $n^* \sim (\ln L)^{1/(k+1)} L^{k/(k+1)}$.
The global density diverges as $\rho \sim L^{2k/(k+1)-s}$

\item $s<k/(k+1)$ --- {\bf Strong High Density Phase} (SHD)

Here $p(n)$ is sharply peaked around $n^* \sim L^{1-s/k}$
and the global density diverges as $n^*$.

\end{itemize}
Numerical evidence for these phases is presented in
Figure~\ref{phasefig} and Figure~\ref{phasefig2}. In
Figure~\ref{phasefig} typical distributions $p(n)$ for the phases
with non-zero density are presented and compare favourably with the
theoretical predictions.
The theoretical predictions were generated by taking the predicted asymptotic forms
of $\lambda$ and inserting into (\ref{ssP}). In the case of Critical Phase A, 
the constant $g$ was taken to be 1
although, $g$ could be used as a parameter to improve the fit to the simulation data.
In Figure~\ref{phasefig2}, $p(n)$ is
plotted in Critical Phase A for different $L$ illustrating the $L$
dependence of the cut-off.

\vspace*{3.0cm}

\subsection{Nearest Neighbour Hopping}
As noted above we expect the mean-field approximation to be
applicable to the fully-connected lattice in the limit $L\to \infty$.
We also carried out numerical simulations of the model
on a one-dimensional lattice with  totally asymmetric hops to nearest neighbour sites  and periodic boundary conditions. The results are given in
Figure~\ref{phasefig} and Figure~\ref{phasefig2} and compare well with the mean-field predictions.

\subsection{The case $b<2$}
As noted in the introduction, in the conserving model with $b<2$
condensation does not occur.
For the non-conserving model the mean-field analysis follows closely that presented above.
One finds that Critical Phase B no longer
exists and 
Critical Phase A becomes a Weak High Density Phase.
Thus for $b<2$ one no longer has critical phases and the phase
diagram reduces to Low Density, Weak High Density and Strong High
Density Phases with phase boundaries given by $s=k$ and $s=k/(k+1)$.

\section{Discussion}
\label{sec:disc}
In this work we have studied a ZRP with
non-conserving dynamics by means of a mean-field theory.
For the  choice of  rates (\ref{uRate}--\ref{aRate})
we find a rich phase diagram with five distinct phases.
For low creation rate we find a low density phase
with exponentially decaying single-site occupation distribution.
On the other hand at high particle creation rate we find a strong
high density phase where the single-site occupation is sharply peaked
at a large occupation which diverges with system size.

The most interesting phases are in the intermediate region.  Here we
find two distinct critical phases and a weak high density phase.  In
the Critical Phase A the occupation distribution decays algebraically
with occupation number with a finite-size cut-off which diverges with
system size.  On the other hand in Critical Phase B there exists a
weak but broad peak at large occupation in addition to the algebraic
decay.  The height and width of this peak scale algebraically with
system size and the area under the peak vanishes.  This peak
corresponds to a large, but subextensive, number of mesocondensates
each containing a large but subextensive number of particles. The
contribution of this peak to the global density vanishes in the
thermodynamic limit leaving the global density as $\rho_c$.  In
addition to these two phase we find a Weak High Density Phase whose
structure is very similar to Critical Phase B except that the
contribution of the weak peak to the global density is dominant and
the global density diverges.

Since the ZRP is a prototypical model for condensation phenomena, we
might expect the phases established here to be displayed in other
driven systems. It would be of interest to explore this possibility by
studying other microscopic models.  For example $p(n)$ given in
(\ref{ssP}) is related to the single-site weight that would be
obtained in a {\em conserving} ZRP with hop rate given by
$\tilde{u}(n) = u(n)+a(n)$.  This would imply a conserving ZRP with
non monotonic hop rates.  Studies of this conserving model have
revealed that the phenomenon of multiple condensates exists there as
well \cite{JSthesis}.

\ack
AGA thanks the Carnegie Trust and US National Science Foundation grant
DMR-0414122 for support.  This study was partially supported by the
Israel Science Foundation (ISF).  Visits of MRE to the Weizmann
Institute were supported by the Albert Einstein Minerva Center for
Theoretical Physics. Visits of DM to Edinburgh were supported by EPSRC
programme grant GR/S10377/01.  We thank the Isaac Newton Institute in
Cambridge, UK for kind hospitality during the programme `Principles of
Dynamics of Nonequilibrium Systems' where part of this project was
carried out.

\end{document}